\documentclass[12pt,fleqn]{article}
\usepackage {multicol}

\usepackage{amssymb,latexsym}
\usepackage{amsmath}

\title{\textbf{Fermi-liquid approach for description of initial stage of fragmentation
at heavy nuclei collisions}}
\author{\textbf{Anatoliy P. Ivashin\footnote{ivashin@kipt.kharkov.ua} ,  Sergey V. Peletminskii, Yuriy V. Slyusarenko \footnote{slusarenko@kipt.kharkov.ua}}\\
 National Science Center "Kharkov Institute of Physics
and Technology"\\
\textit{Akademicheskaya Str.,1, Kharkov, 61108, Ukraine}}

\date{}

\begin{document}

 \maketitle
\begin{flushleft}
UDK: 530.145
\end{flushleft}

\begin{abstract}
A mechanism is proposed for initial stage of instability development
that can induce the
fragmentation of nuclear matter, arising as a result of collisions of
non-relativistic heavy nuclei.
Collision of heavy nuclei is simulated
as a collision of two
unbounded Fermi-liquid ``drops''.

The instability
origination in such a system is related to propagation of increasing oscillations in
the nuclear matter.
These oscillations can
exist in a resting Fermi-liquid: modified Landau zero sound,
modified spin and isospin waves, combination of these more simple
waves. These instabilities are analogous to the beam instability
in ordinary electron plasma. Behavior features of the obtained
oscillation increase increments are provided.
They can be used as
indication for experimental confirmation of the proposed mechanism
of  fragmentation at nuclear collisions. Directions along which nuclear
matter ``jets'' can be expected are specified. 
\end{abstract}


\begin{multicols}{2}

\section{Introduction}


The fragmentation phenomenon, i. e. the simultaneous decay of the excited
nucleus into lighter nuclei (fragments) and separate particles, has been known
for a rather long time. However the scientific interest to this phenomenon
have been rapidly growing lately due to the experiments on collisions of
heavy fast nuclei, which were carried out in different scientific centers
of the world (CERN, Switzerland; Dubna, Russia; Berkley, Oak Ridge, USA;
Hamburg, Germany; Orsay, Caen, France).
This interest is caused by the prospect to confirm our notions about
inner structure of nuclei and the character of inter-nucleon interactions
and also by the possibility to obtain new fundamental knowledge about
the origin of intranuclear and intranucleon forces.

Nuclear matter formed by the collision of heavy nuclei
 is an unstable object decaying within a very short time.
The decay of arisen nuclear matter is accompanied by the forming of new
lighter nuclei and nuclear fragments (so-called nuclear fragmentation
process). Nowadays while describing the evolution of nuclear matter,
formed after heavy nuclei collision, one usually considers two scenarios
(see for example \cite{jR,bBo}). According to the first
one the state of statistical equilibrium can be reached in the nuclear
matter (the thermalization of the nuclear matter) permitting the
thermodynamic description of the fragmentation process by means of the
phase transition theory methods. According to the second scenario the
resulting nuclear matter is essentially unstable matter, in which the
processes allow only the dynamic description.

In the present paper we assume the latter scenario. Each colliding
nucleus is a system of nucleons and it can be considered as a generalized
Fermi-liquid, whose particles have spin and isospin (distinguishing
protons and neutrons) degrees of freedom. Let us note that the
Fermi-liquid approach of Landau and Silin \cite{lD,vS1} is widely
applied in theoretical nuclear physics in order to describe the
properties of heavy nuclei \cite{aM1,bB}. In solving the specific
problems of theoretical nuclear physics it is often assumed that
heavy nuclei (with $A>100$) can be considered as the infinite drops
of nucleonic Fermi liquid in cases when surface effects are unimportant.
Then such assumption enables us to simplify complicated mathematical
treatment without violation of the final results. Note, that theoretical
approach for finite Fermi liquids was elaborated by A.B.Migdal ( see
 \cite{aM1} and references therein).

In the present paper we also suppose the nuclear matter, formed due to
heavy nuclei collisions, to be the Fermi-liquid consisting of  particles
with spin and isospin degrees of freedom. The inter-particle interaction
in this kind of nuclear liquid is described by such parameters as Landau
amplitudes, by analogy with the normal Fermi-liquid theory \cite{lD,vS1}.
So in order to consider the fragmentation formed after collision we come
to the less complicated problem of interaction of two drops of nucleon
infinite Fermi-liquid moving relatively to each other. Note that description
of collisions of heavy ions as two drops of infinite quark Fermi-liquids
leading to formation of quark-gluon plasma was analyzed in paper \cite{aI1}.

We have demonstrated that interactions of two nucleon Fermi-liquids
in the system can originate a number of instabilities associated with
increased oscillations of modes existing in the ordinary Fermi-liquid.
Such waves include a modified Landau zero sound, modified spin waves,
waves related to excitation of isospin degrees of freedom, as well as
more complex waves related to a combination of the mentioned simple
waves \cite{bB}.The present instabilities are analogous to the known
"beam'' instability in the ordinary electron plasma, occurring when
charged particle beams pass through the plasma \cite{aI2,dB}.
Since the mentioned instabilities can evolve exactly after
the mutual penetration of colliding Fermi-liquid drops and formation
of united Fermi-liquid, we consider the formed system as an object
which is far from the statistical equilibrium and requiring the dynamical
description.
The dynamics of such system in this work is described by the kinetic
equation for the one-particle distribution function for nucleons in
collisionless approximation, which is widely used in the theory of
ordinary Fermi-liquid \cite{lD,vS1}. Such approach for phenomena
we consider is valid in the case when the characteristic time of
instability evolution is less in comparison with the characteristic
time of relaxation in our system. In many cases such approach allows
the analytic description of the initial stage of instability development
in formed nuclear matter without numerical calculations. Note that the
use of mentioned kinetic equation restricts our consideration to
non-relativistic nucleon energies. The modification of the kinetic
equation for the description of evolution of the nuclear matter
with relativistic nucleons is given in \cite{yuI,yuI1}. In these papers
the development of filamentary instabilities in the nuclear matter
at the stage of the mutual penetration of the two colliding nuclei
was considered on the basis of the kinetic equation for relativistic
nucleons.

Use of this approach allows us to describe the initial stage of
instability development in nuclear matter, formed after the nuclear
collision and to obtain the dispersion equations for zero sound
oscillations in such system (analogue of zero sound oscillations in
ordinary Fermi-liquid). The solutions of equations for small Landau
amplitudes are investigated in details in case of collision of
nuclei in normal state as well as in case of collisions of heavy
excited nuclei. The expressions for the increments of zero-sound
oscillations in nuclear matter are found. We have shown that those
increments have characteristic logarithmic dependence on the excitation
energy of  nucleus when the nuclear matter is formed due to the
collision of heavy excited nuclei. This characteristic dependence
can be used for experimental proof or disproof of the mechanism
of the initiation of instabilities in nuclear matter which is
suggested in present paper.

We predict the directions (angular distribution) in which it
is possible to expect the matter jets, consisting of the nuclei
fragments and separate nucleons and originating as a result of
the destruction of nuclei under collision. In this paper these
propagation directions are related to the direction and value
of the relative velocity of two colliding nuclei. The occurrence
of such jets is caused in our opinion by the further evolution
 of instabilities, described in this paper. It is known that the
 experimental exploration of the angular distributions of outgoing
matter jets is the important data source revealing the processes
in formed dense blob of nuclear matter. For this purposes the
special experimental complexes are created, they allow to perform
the mentioned measurements almost in the whole spatial angle $4\pi$
\cite{bBo,jP}. The experimental detection
of mentioned jets in predicted directions would by evidence of
the realization of suggested mechanism of initiation of instabilities
in colliding nuclei system.

We suppose the confirmation of predictions given in present paper should
be expected for example in experiments on nuclear collisions Gd+U or Xe+Sn
(INDRA, \cite{bBo,jP}) with incoming nucleus energy
above 145 MeV per nucleon.

\section{ Basic kinetic equations of two colliding Fermi-liquids}

A solution of the formulated problem on kinetics of two nucleon
Fermi-liquids collisions is based on the assumption that the Fermi-liquid
energy is a functional $E\left( {f} \right)$ of one-particle
distribution function $f_{i} \left( {\mathbf{x},\mathbf{p},t}
\right)$ of quasiparticles (nucleons; here index ``\textit{i}'' is
used to denote a quasiparticle spin and isospin, $ {\mathbf{x}} $
is a coordinate and $\mathbf{p}$ is a quasiparticle momentum,
\textit{t} is a time). We want to emphasize that in the
Fermi-liquid theory the introduced energy functional plays the
role analogous to Hamiltonian in the microscopic theory.

Evolution of one-particle distribution function $f_{i} \left( {\mathbf{x},\mathbf{p},t}
\right)$ in the collisionless approximation ($\omega \tau _{r} \gg 1$,
$\omega $ is a frequency, $\tau _{r}$ is a relaxation time) is governed by the
kinetic equation

\begin{equation}
\label{eq1} \frac{{\partial f_{i}} }{{\partial t}} +
\frac{{\partial \varepsilon _{i} }}{{\partial
\mathbf{p}}}\frac{{\partial f_{i}} }{{\partial \mathbf{x}}} -
\frac{{\partial \varepsilon _{i}} } {{\partial
\mathbf{x}}}\frac{{\partial f_{i}} }{{\partial \mathbf{p}}} = 0,
\end{equation}

\noindent where quasiparticle energy $\varepsilon _{i} \left(
{\mathbf{x},\mathbf{p},f} \right)$, determining kinetics of the
nucleon system, represents a variational derivative of the energy
functional over the one-particle distribution function

\[
\varepsilon _{i} \left( {\mathbf{x},\mathbf{p},f} \right) = \frac{{\delta  E \left( {f}
\right)}}{{\delta f_{i} \left( {\mathbf{x},\mathbf{p}} \right)}}.
\]

A specific form of quasiparticle energy $\varepsilon _{i} \left(
{\mathbf{x},\mathbf{p},f} \right)$ as a functional of the
distribution function in the Fermi-liquid theory is not known. Therefore,
let us introduce functions of quasiparticle interaction
\textit{F}$_{ij} $ (generalized Landau amplitudes), which
represent a linear reaction of the quasiparticle energy on small
deviation $\delta f_{i} \left( {\mathbf{x},\mathbf{p},t} \right)$
of distribution function~$f_{i} \left( {\mathbf{x},\mathbf{p},t}
\right)$:

\noindent
\begin{gather}
\delta \varepsilon _{i} \left( {\mathbf{x},\mathbf{p},t} \right)=
{\text{\hfill}}
\nonumber
\\
= \sum\limits_{j} \int d\tau ^{'}\int d\mathbf{x}^{'}F_{ij}
\left( {\mathbf{x} -
\mathbf{x}^{'};\mathbf{p},\mathbf{p}^{'}} \right)\times
\nonumber
 \\
\times \delta f_{j} \left( {\mathbf{x}^{'},\mathbf{p}^{'},t}
\right)  ,\label{eq2}
\end{gather}

\noindent where $d\tau = d^{3}p/\left( {2\pi \hbar}  \right)^{3}$.
Functions \textit{F}$_{ij}$, representing second order  variational
derivatives of the energy functional with respect to the
one-particle distribution function, are main characteristics of the
theory that can be experimentally determined (in this connection
see, for example \cite{sB1,sB2}). In view of short-range
interaction forces between nucleons, from here on we will consider
that the functions \textit{F}$_{ij}$ have the following form:

\begin{equation}
\label{eq3}
F_{ij} \left( {\mathbf{x} - \mathbf{x}^{'},\mathbf{p},\mathbf{p}^{'}} \right) =
 F_{ij} \left( {\mathbf{p},\mathbf{p}^{'}}
\right)\delta \left( {\mathbf{x} - \mathbf{x}^{'}} \right),
\end{equation}

\noindent
where quantities $F_{ij} \left( {\mathbf{p},\mathbf{p}^{'}} \right)$
(we will call them also Landau
amplitudes) do not depend on coordinates. Note that
assumption (\ref{eq3}) is valid only if we neglect the Coulomb forces induced by the
proton charge. Linearizing equation (\ref{eq1}) near a stationary quasi-equilibrium
state (see below), described by distribution function $f_{0i} $, we can go
over to Fourier components of deviations $\delta f_{i} = f_{i} - f_{0} $ of
the one-particle distribution functions from their equilibrium values

\begin{gather}
 \widetilde{\delta f}_{i} \left( {\mathbf{p},\omega
,\mathbf{k}} \right) =\nonumber \\
= \int {d^{3}x\int\limits_{ - \infty}
^{\infty}  {dt} \delta f_{i} \left( {\mathbf{x},\mathbf{p},t}
\right)\exp\left( {i\omega t - i\mathbf{k}\mathbf{x}} \right)}
\label{eq4}
\end{gather}

As a result, using (\ref{eq2}) we get the following kinetic equation for
${\widetilde{\delta f}}_{i} \left(
{\mathbf{p},\omega ,\mathbf{k}} \right)$:

\begin{gather}
\left( {\omega - \mathbf{k}\frac{{\partial \varepsilon _{0i}} }{{\partial \mathbf{p}}}}
\right)\widetilde{\delta f}_{i} ( {\mathbf{p},\omega ,\mathbf{k}} ) +\nonumber\\
+\mathbf{k}\dfrac{{\partial f_{0i}} }{{\partial \mathbf{p}}}\int d\tau ^{'}\sum\limits_{j}
{F_{ij} \left( {\mathbf{p},\mathbf{p}^{'}} \right)}
\widetilde{\delta f}_{i} ({\mathbf{p}^{'},\omega ,\mathbf{k}} ) = 0,
\label{eq5}
\end{gather}

\noindent
where amplitudes $F_{ij} \left( {\mathbf{p},\mathbf{p}^{'}} \right)$
in accordance with (\ref{eq2}),
(\ref{eq3}) are given by the formula

\[
F_{ij} \left( {\mathbf{p},\mathbf{p}^{'}} \right) = \left. {\frac{{\delta ^{2}E\left( {f}
\right)}}{{\delta f_{i} \left( {\mathbf{p}} \right)\delta f_{j} \left( {\mathbf{p}^{'}}
\right)}}} \right|_{f_{i} = f_{i0}}
\]

\noindent
and $\varepsilon _{0i} $  is the energy of noninteracting nucleons, $\varepsilon
_{0} = \dfrac{p^{2}}{2m}$ (\textit{m} is the nucleon mass).

If we assume that interaction between nucleons is invariant with respect to
the spin and isospin transformations, then a structure of Landau amplitudes
\textit{F} for the nucleon liquid will be of the following form:

\begin{gather}
F =  I^{\left( {s} \right)}I^{\left( {i} \right)}F^{\left( {0} \right)}
+ I^{\left( {i} \right)}\left(\boldsymbol {\sigma \sigma}  \right)F^{\left( {s}
\right)} +\nonumber\\
+ I^{\left( {s} \right)}\left( \boldsymbol{\tau \tau}  \right)F^{\left( {i}
\right)}
 + \left( \boldsymbol {\sigma  \sigma}  \right)\left( \boldsymbol{\tau \tau}
\right)F^{\left( {si} \right)}
\label{eq6}
\end{gather}

\noindent
where $F^{\left( {0} \right)},F^{\left( {s} \right)},F^{\left( {i}
\right)},F^{\left( {s,i} \right)}$ are generalized Landau amplitudes,
$I^{\left( {s} \right)},\boldsymbol{\sigma} $ are unit matrix and Pauli matrices in the
spin space,
 $I^{\left( {i} \right)},\boldsymbol{\tau} $ are a unit matrix and Pauli
matrices in the isospin space. Accordingly in equations (\ref{eq5}) there will be one
of the Landau amplitudes from expression (\ref{eq6}), depending on which
oscillations are considered (oscillations of density, spin density, isospin
density or spin-isospin density). Thereby, hereinafter we will deal with
only one linearized kinetic equation

\begin{gather}
\left( {\omega - \mathbf{k}\frac{{\partial \varepsilon _{0}}
}{{\partial \mathbf{p}}}} \right) \widetilde{\delta f}(
{\mathbf{p},\omega ,\mathbf{k}} ) +\nonumber\\
+\mathbf{k} \frac{{\partial
f_{0}} }{{\partial \mathbf{p}}}
\int d\tau ^{'}F\left( {\mathbf{p},\mathbf{p}^{'}}\right)
\widetilde{\delta f}( {\mathbf{p},\omega ,\mathbf{k}} ) = 0,
\label{eq7}
\end{gather}

\noindent
where we will bear in mind $F\left( {\mathbf{p},\mathbf{p}^{'}} \right)$ as one of the
amplitudes from~(\ref{eq6}).

Now let us obtain the expression for a quasi-equilibrium distribution
function from kinetic equation (\ref{eq7}). The equilibrium
distribution functions of the resting drops of Fermi-liquid
will be denoted by $f_{0}^{\left( {1} \right)} \left( {\mathbf{p}}
\right)$ and $f_{0}^{\left( {2} \right)} \left( {\mathbf{p}}
\right)$. Then in the case when the first drop is resting and the
second one
(impacting) is moving with velocity $\mathbf{u}$,
equilibrium distribution functions will be $f_{0}^{\left( {1}
\right)} \left( {\mathbf{p}} \right)$ and $f_{0}^{\left( {2}
\right)} \left( {\mathbf{p} - m\mathbf{u}} \right)$, respectively.
This implies that the quasi-equilibrium distribution function of
two colliding unbounded Fermi-liquids consisting of identical
particles has the form

\begin{equation}
\label{eq8}
f_{0}^{} \left( {\mathbf{p}} \right) = \alpha _{1} f_{0}^{\left( {1} \right)}
\left( {\mathbf{p}} \right) + \alpha _{2} f_{0}^{\left( {2} \right)} \left( {\mathbf{p} - m\mathbf{u}}
\right).
\end{equation}

\noindent
 Obviously that this function obeys the kinetic equation~(\ref{eq1}).

We want to explain a meaning of coefficients $\alpha _{1} $ and
$\alpha _{2} $ appearing in this expression. With this aim we will introduce
auxiliary distribution functions $g_{0}^{} \left( {\mathbf{p}} \right)$,
$g_{0}^{\left( {1} \right)} \left( {\mathbf{p}} \right)$, $g_{0}^{\left( {2} \right)}
\left( {\mathbf{p}} \right)$ normalized to unity (and because of that  permitting
a probability interpretation) and related to each other analogously to relation~(\ref{eq8}):

\begin{equation}
\label{eq9}
g_{0}^{} \left( {\mathbf{p}} \right) = q_{1} g_{0}^{\left( {1} \right)} \left(
{\mathbf{p}} \right) + q_{2} g_{0}^{\left( {2} \right)} \left( {\mathbf{p} - m\mathbf{u}} \right).
\end{equation}

From the normalization condition of these functions to unit it follows that values
\textit{q}$_{1}$ and\textit{ q}$_{2} $ satisfy the equality

\begin{equation}
\label{eq10}
q_{1} + q_{2} = 1.
\end{equation}

From here we see that coefficients $q_{1} $ and $q_{2} $ must be interpreted
as probabilities that
an arbitrary selected particle belongs to either a system
with distribution function $g_{0}^{\left( {1} \right)} \left( {\mathbf{p}} \right)$,
or to a system with the distribution function $g_{0}^{\left( {2} \right)}
\left( {\mathbf{p}} \right)$. However, in this case probabilities $q_{1} $
and\textit{} $q_{2} $ must be specified by relative frequencies of the
events of finding objects
with distribution functions $g_{0}^{\left(
{1} \right)} \left( {\mathbf{p}} \right)$ and $g_{0}^{\left( {2} \right)} \left( {\mathbf{p}}
\right)$. If we are dealing with two colliding beams of nuclei, then for
probabilities $q_{1} $ and\textit{} $q_{2} $, we can obtain the following
expressions (see (\ref{eq10}))

\begin{equation}
\label{eq11}
q_{1} = N_{1} /\left( {N_{1} + N_{2}}  \right),
\quad
q_{2} = N_{2} /\left( {N_{1} + N_{2}}  \right),
\end{equation}

\noindent
where $N_{1} $ and $N_{2} $ are densities of nuclei in the first and second
beams respectively.

As it was previously noted the above given argumentation implies
the normalization of the distribution functions to unity. In the
case under consideration this requirement is not fulfilled for
functions $f_{0}^{} \left( {\mathbf{p}} \right)$, $f_{0}^{\left(
{1} \right)} \left( {\mathbf{p}} \right)$, $f_{0}^{\left( {2}
\right)} \left( {\mathbf{p}} \right)$ because

\begin{gather}
\int {d\tau}  f_{0}^{} \left( {\mathbf{p}} \right) = n,
\,
\int {d\tau}  f_{0}^{\left( {1} \right)} \left( {\mathbf{p}} \right) = n_{1} ,\nonumber\\
\int {d\tau}  f_{0}^{\left( {2} \right)} \left( {\mathbf{p}} \right) = n_{2} ,
\label{eq12}
\end{gather}

\noindent where \textit{n} is density of number of particles in
the system consisting of two colliding Fermi-liquids; $n_{1}  $
and $n_{2} $ are densities of number of particles in each of these
Fermi-liquids. But assuming in (\ref{eq9}) that

\begin{gather}
g_{0}^{} \left( {\mathbf{p}} \right) = \frac{{f_{0} \left( {\mathbf{p}} \right)}}{{n}},
\,
g_{0}^{\left( {1} \right)} \left( {\mathbf{p}} \right) = \frac{{f_{0}^{\left( {1}
\right)} \left( {\mathbf{p}} \right)}}{{n_{1}} },\nonumber\\
g_{0}^{\left( {2} \right)} \left( {\mathbf{p}} \right) = \frac{{f_{0}^{\left( {2}
\right)} \left( {\mathbf{p}} \right)}}{{n_{2}} }\nonumber
\end{gather}

\noindent
and introducing denotations

\begin{equation}
\label{eq13}
\alpha _{1} = nq_{1} /n_{1} ,
\quad
\alpha _{2} = nq_{2} /n_{2} ,
\end{equation}

\noindent
we can arrive at the expression~(\ref{eq8}).

\section{Zero-sound dispersion equation in a simple model}

In order to solve kinetic equation~(\ref{eq7}) we will use a simple model, where the
Landau amplitudes do not depend on momenta $\mathbf{p}$, $\mathbf{p}^{'}$.
We also assume that
the colliding Fermi-liquids have zero temperature, so that the equilibrium
distribution functions of separate drops are given by the formulas

\begin{gather}
f_{0}^{\left( {1} \right)} \left( {\mathbf{p}} \right) = \theta \left( {\varepsilon
_{1F} - \varepsilon}  \right),
f_{0}^{\left( {2} \right)} \left( {\mathbf{p}} \right) = \theta \left( {\varepsilon
_{2F} - \varepsilon}  \right),\nonumber\\
\varepsilon = \dfrac{{p^{2}}}{{2m}} = \varepsilon _{0} \left( {\mathbf{p}} \right),
\label{eq14}
\end{gather}

\noindent where $\theta $($\varepsilon $) is theta function,
$\varepsilon _{1F} $, $\varepsilon _{2F} $ are Fermi energies of
the colliding drops. It is evident from (\ref{eq14}) that in this
case a difference between nucleus types can be found only in their
different Fermi energies (note that for heavy nuclei $\varepsilon
_{1F} \approx \varepsilon_{2F}$).

Within used assumptions linearized kinetic equation (\ref{eq7}) can be written
as:

\begin{gather}
\widetilde{\delta f}( {\mathbf{p},\omega ,\mathbf{k}} )
\left( {\omega - \mathbf{k}\mathbf{v}} \right) +\nonumber\\
+\mathbf{k}\frac{{\partial f_{0} \left( {\mathbf{p}} \right)}}
{{\partial \mathbf{p}}}F\int {d\tau}
\widetilde{\delta f}( {\mathbf{p},\omega ,\mathbf{k}} ) = 0,
\label{eq15}
\end{gather}

\noindent
where $\mathbf{v} = \dfrac{{\partial \varepsilon} }{{\partial \mathbf{p}}} =
\dfrac{{\mathbf{p}}}{{m}}$,
and in accordance with (\ref{eq8}), (\ref{eq14})

\begin{eqnarray}
\label{eq16}
\frac{{\partial f_{0}} }{{\partial \mathbf{p}}} &=& - \alpha _{1} \mathbf{v}\delta
\left(
{\varepsilon _{1F} - \varepsilon}  \right) - \alpha _{2} \left( {\mathbf{v} -
\mathbf{u}}
\right)\times\nonumber\\
&\times&\delta \left( {\varepsilon _{2F} - \dfrac{{m\left( {\mathbf{v} - \mathbf{u}}
\right)^{2}}}{{2}}} \right).
\end{eqnarray}

The most general solution of equation (\ref{eq15}) looks as

\begin{gather}
\widetilde {\delta f}\left( {\mathbf{p},\mathbf{k},\omega}  \right) =\nonumber\\
= - F\left\{ {\omega -
 \mathbf{k}\mathbf{v} +
io} \right\}^{-1}\mathbf{k}\frac{{\partial f_{0} \left( {\mathbf{p}} \right)}}{{\partial
\mathbf{p}}}\delta \varphi \left( {\omega ,\mathbf{k}} \right) + \nonumber\\
+ \delta A\left( {\mathbf{p},\mathbf{k}}
\right)\delta \left( {\omega - \mathbf{v}\mathbf{k}} \right),
\label{eq17}
\end{gather}

\noindent
where we denoted

\begin{equation}
\label{eq18}
\delta \varphi \left( {\omega ,\mathbf{k}} \right) \equiv \int
{d\tau \widetilde {\delta f}\left( {\mathbf{p},\mathbf{k},\omega}  \right)} ,
\end{equation}

\noindent
and $\delta A\left( {\mathbf{p},\mathbf{k}} \right)$ are arbitrary functions
satisfying the
condition (see (\ref{eq4}))

\begin{equation}
\label{eq19}
\delta A^{\ast} \left( {\mathbf{p},\mathbf{k}} \right) = \delta A\left( {\mathbf{p}, -
 \mathbf{k}} \right).
\end{equation}

\noindent
Formula (\ref{eq17}) permits to find a value of $\delta \varphi \left( {\omega ,\mathbf{k}}
\right)$ in terms of the functions $\delta A\left( {\mathbf{p},\mathbf{k}} \right)$:

\begin{equation}
\label{eq20}
\delta \varphi \left( {\omega ,\mathbf{k}} \right) = \tilde {\varepsilon} ^{ -
1}\left( {\omega ,\mathbf{k}} \right) \widetilde {\delta A}\left( {\omega ,\mathbf{k}} \right),
\end{equation}

\noindent
where (see(\ref{eq19}))

\begin{equation}
\label{eq21}
\widetilde {\delta A}\left( {\omega ,\mathbf{k}} \right) = \int {d\tau \delta A\left(
{\mathbf{p},\mathbf{k}} \right)\delta \left( {\omega - \mathbf{k}\mathbf{v}} \right)} ,
\end{equation}

\[
\widetilde {\delta A}^{*}\left( {\omega ,\mathbf{k}} \right) = \widetilde {\delta A}
\left( {-\omega ,-\mathbf{k}} \right).
\]

Substituting further (\ref{eq20}) into (\ref{eq17}), we will have the following relation for
$\widetilde {\delta f}\left( {\mathbf{p},\mathbf{k},\omega}  \right)$:

\begin{gather}
\widetilde {\delta f}\left( {\mathbf{p},\mathbf{k},\omega}  \right) =\nonumber\\
= \delta A
\left( {\mathbf{p},\mathbf{k}}
\right)\delta \left( {\omega - \mathbf{k}\mathbf{v}} \right)
- F\left\{ {\omega - \mathbf{k}\mathbf{v} + io}
\right\}^{ - 1}\times  \nonumber\\
\times\tilde {\varepsilon} ^{ - 1}\left( {\omega ,\mathbf{k}} \right)
\widetilde {\delta A}\left( {\omega ,\mathbf{k}} \right)\mathbf{k}\frac{{\partial f_{0}
\left( {\mathbf{p}}
\right)}}{{\partial \mathbf{p}}},
\label{eq22}
\end{gather}

\noindent
where $\tilde {\varepsilon} \left( {\omega ,\mathbf{k}} \right)$ in (\ref{eq20}),
(\ref{eq22}) is
defined by the formula

\begin{gather}
\tilde {\varepsilon} \left( {\omega ,\mathbf{k}} \right) = \tilde {\varepsilon} ^{ *
}\left( { - \omega , - \mathbf{k}} \right) = \tilde {\varepsilon} _{1} \left( {\omega
,\mathbf{k}} \right) + i\tilde {\varepsilon} _{2} \left( {\omega ,\mathbf{k}} \right)\nonumber\\
 = 1 +
F\mathbf{k}\int {d\tau}  \frac{{\partial f_{0} \left( {\mathbf{p}} \right)}}{{\partial
\mathbf{p}}}\left\{ {\omega - \mathbf{k}\mathbf{v} + io} \right\}^{ - 1}_.
\label{eq23}
\end{gather}

For the case of the charged Fermi-liquid $\tilde {\varepsilon} $
represents a complex dielectric permittivity (see, for example,
\cite{aI3}). As is known the presence of an imaginary additonal
component in $\tilde {\varepsilon} \left( {\omega ,\mathbf{k}} \right)$
indicates damping or increase of wave amplitudes. The wave
dispersion law $\omega \left( {k} \right) = \omega _{0} \left( {\mathbf{k}}
\right) + i\gamma \left( {\mathbf{k}} \right)$ should be found from the
equation

\begin{equation}
\label{eq24}
\tilde {\varepsilon} \left( {\omega _{0} \left( {\mathbf{k}} \right) + i\gamma \left(
{\mathbf{k}} \right),\mathbf{k}} \right) = 0,
\end{equation}

\noindent
with decrement (increment) $\gamma \left( {\mathbf{k}} \right)$ determined by the
imaginary part of quantity $\tilde {\varepsilon} \left( {\omega ,\mathbf{k}}
\right)$. This is a reason why weakly damping or weakly increasing
oscillations in the system

\begin{equation}
\label{eq25}
\left| {\omega _{0} \left( {\mathbf{k}} \right)} \right| \gg \left| {\gamma \left(
{\mathbf{k}} \right)} \right|
\end{equation}

\noindent
can exist only on condition that

\begin{equation}
\label{eq26}
\left| {\tilde {\varepsilon} _{1} \left( {\omega ,\mathbf{k}} \right)} \right| \gg
\left| {\tilde {\varepsilon} _{2} \left( {\omega ,\mathbf{k}} \right)} \right|.
\end{equation}

\noindent
Using the formula

\[
\left( {z + io} \right)^{ - 1} = P\frac{{1}}{{z}} - i\pi \delta \left( {z}
\right)
\]

\noindent
(where \textit{P} is a symbol of the principal value), and taking into
account (\ref{eq16}), we will write functions $\tilde {\varepsilon} _{1} \left(
{\omega ,k} \right)$, $\tilde {\varepsilon} _{2} \left( {\omega ,k} \right)$
(see (\ref{eq23})) in the following form:

\end{multicols}

\begin{equation*}
\tilde {\varepsilon} _{1} \left( {\omega ,k} \right) = 1 + \alpha
_{1} {\cal F}\left\{ {1 - \frac{{\omega} }{{2kv_{1F}} }\ln\left|
{\frac{{\omega + kv_{1F} }}{{\omega - kv_{1F}} }} \right|}
\right\}
 +  \frac{\alpha _{2}{v_{2F}}{\cal F} }{{v_{1F}} }\left\{ {1 -
\frac{{\omega - ku\cos\alpha} }{2{kv_{2F}} }\ln\left|
{\frac{{\omega - ku\cos\alpha + kv_{2F}} }{{\omega - ku\cos\alpha - kv_{2F}
}}} \right|} \right\},
\end{equation*}

\begin{equation}
\tilde {\varepsilon} _{2} \left( {\omega ,k} \right) = \frac{{\pi
}}{{2}}{\cal F}\left\{ {\alpha _{1} \frac{{\omega} }{{kv_{1F}} }\theta \left( {1 -
\left| {\frac{{\omega} }{{kv_{1F}} }} \right|} \right) + \alpha _{2}
\left(\frac{{v_{2F}} }{{v_{1F}} }\right)
\frac{{\omega - ku\cos\alpha} }{{kv_{2F}} }\theta
\left( {1 - \left| {\frac{{\omega - ku\cos\alpha} }{{kv_{2F}} }} \right|}
\right)} \right\}.
\label{eq27}
\end{equation}

\begin{multicols}{2}
\noindent
Here ${\cal F}$ is a dimensionless Landau amplitude

\begin{equation}
\label{eq28}
{\cal F} \equiv \frac{{Fm^{2}v_{1F}} }{{2\pi ^{2}\hbar ^{3}}},
\end{equation}

\noindent
velocities $v_{1F} ,v_{2F} $ are given by relations

\begin{equation}
\label{eq29}
\varepsilon _{1F} = \frac{{mv_{1F}^{2}} }{{2}},
\quad
\varepsilon _{2F} = \frac{{mv_{2F}^{2}} }{{2}}
\end{equation}

\noindent and angle $\alpha $ is an angle between directions of
wave vector $\mathbf{k}$ and velocity of impacting
drop $\mathbf{u}$.
In accordance with (\ref{eq24}) - (\ref{eq27}) a dispersion equation has the form
$\tilde {\varepsilon} _{1}
\left( {\omega
_{0} \left( {k} \right),k} \right) = 0$, or

\begin{gather}
1 + \alpha _{1} {\cal F}\left\{ {1 - \frac{{s}}{{2}}\ln\left| {\frac{{s + 1}}{{s -
1}}} \right|} \right\} + \alpha _{2} \frac{{1}}{{\eta} }{\cal F}\times\nonumber\\
\times\left\{ {1 -
\frac{{1}}{{2}}\left( {\eta s - s_{0}}  \right)\ln\left| {\frac{{\eta s -
s_{0} + 1}}{{\eta s - s_{0} - 1}}} \right|} \right\} = 0,
\label{eq30}
\end{gather}

\noindent
and the decrement (increment) is given by

\begin{equation}
\label{eq31}
\gamma \left( {k} \right) = \left\{ {\frac{{\partial \tilde {\varepsilon
}_{1} \left( {\omega ,k} \right)}}{{\partial \omega} }} \right\}_{\omega =
\omega _{0}} ^{ - 1} \tilde {\varepsilon} _{2} \left( {\omega ,k} \right),
\end{equation}

\begin{gather*}
\tilde {\varepsilon} _{2} \left( {\omega ,k} \right)
= \dfrac{{\pi
}}{{2}}{\cal F}
\left\{ \alpha _{1} s\theta \left( {1 - \left| {s} \right|} \right)+\right.\\
+\left. \alpha _{2} \frac{{1}}{{\eta} }\left( {\eta s - s_{0}}  \right)\theta
\left( {1 - \left| {\eta s - s_{0}}  \right|} \right) \right\}.
\end{gather*}

We use the following denotations in formulas (\ref{eq30}), (\ref{eq31}) (see also (\ref{eq28}),
(\ref{eq29})):

\begin{equation}
\label{eq32}
s \equiv \frac{{\omega _{0}} }{{kv_{1F}} },
\quad
\eta \equiv \frac{{v_{1F}} }{{v_{2F}} },
\quad
s_{0} \equiv \frac{{u}}{{v_{2F}} }\cos\alpha
\end{equation}

\noindent
(remember that for the heavy nuclei $\varepsilon _{1F} \approx \varepsilon
_{2F} $, that is, $\eta \approx 1$).

The derivative $\left\{ {\dfrac{{\partial \tilde {\varepsilon} _{1} \left(
{\omega ,k} \right)}}{{\partial \omega} }} \right\}_{\omega = \omega _{0}}
$, in (\ref{eq31}), according to (\ref{eq27}), (\ref{eq30}), acquires the form

\end{multicols}

\begin{equation}
\label{eq33}
\left\{ {\frac{{\partial \tilde {\varepsilon} _{1} \left( {\omega
,k} \right)}}{{\partial \omega} }} \right\}_{\omega = \omega _{0}}
= \frac{{s}}{{\omega _{0}} }\left\{ { - \frac{{\alpha _{1} {\cal
F}}}{{2}}\ln\left| {\frac{{s + 1}}{{s - 1}}} \right| }
 { + \frac{{\alpha _{1} {\cal F}s}}{{s^{2} - 1}} - \frac{{\alpha _{2}
{\cal F}}}{{2}}\ln\left| {\frac{{\eta s - s_{0} + 1}}{{\eta s - s_{0} - 1}}} \right|
+ \alpha _{2} {\cal F}\frac{{\eta s - s_{0}} }{{\left( {\eta s - s_{0}}
\right)^{2} - 1}}} \right\}.
\end{equation}
\begin{multicols}{2}

\section{Solution of the dispersion equation for zero-sound
oscillations}

A dispersion law of the zero-sound oscillations $\omega = \omega
_{0} \left( {k} \right)$ is determined by the solution of equation
(\ref{eq30}). Note first of all that this equation (in the same
way as the relation for $\gamma \left( {k} \right)$, see
(\ref{eq31}), (\ref{eq33})) is invariant with respect to a
simultaneous substitution $s \to - s,\; s_{0} \to - s_{0} $, (see
(\ref{eq32})). This condition shows that propagation of the
zero-sound wave is possible in opposite directions with the same
increment or decrement. Consequently, let us consider $s > 0$ for
the sake of definiteness. Since equation (\ref{eq30}) is quite
complicated in its general form, we will make certain assumptions
in order to obtain a solution.

It is known that in the ordinary Fermi-liquid propagation of
undamped zero-sound oscillations at zero temperature is possible
only under the condition ${\cal F} > 0,\quad s > 1$ (in this
connection see \cite{dP,pP}). Otherwise oscillations will quickly
damp. Let us make an assumption that

\begin{equation}
\label{eq34} {\cal F} > 0,\quad s > 1
\end{equation}

\noindent and demonstrate that in this case oscillations can
increase or damp due to existence of the impacting drop.
Naturally, one can easily see from (\ref{eq31}) that possibility
of propagation of weakly increased or weakly damped oscillations
in the case, determined by inequalities (\ref{eq34}), is related
to the fulfilment of the condition

\begin{equation}
\label{eq35} \left| {\eta s - s_{0}}  \right| < 1.
\end{equation}

\noindent If inequalities (\ref{eq34}), (\ref{eq35}) are valid,
quantity $\tilde {\varepsilon} _{2} \left( {\omega _{0} ,k}
\right)~\equiv~\tilde {\varepsilon} _{2} \left( {s} \right)$ has
the form:

\begin{equation}
\label{eq36} \tilde {\varepsilon} _{2} \left( {s} \right) = \alpha
_{2} \frac{{\pi }}{{2\eta} }\left( {\eta s - s_{0}}  \right){\cal
F}.
\end{equation}

Let us study the solution of dispersion equation (\ref{eq30}) with
small Landau amplitudes ${\cal F}$, ${{\cal F} \ll 1}$ and at the
conditions (\ref{eq34}), (\ref{eq35}).
 Since in this case quantity
\textit{s} is close to unity, $s~\approx~1~+~\delta s$, equation
(\ref{eq30}) can be written as

\begin{gather*}
1 + \alpha _{1} {\cal F} + \frac{{\alpha _{2}} }{{\eta} }{\cal
F} - \frac{{\alpha _{2} }}{{2\eta} }{\cal F}\left( {\eta - s_{0}}
\right)\times\\
\times\ln\left\{ {\frac{{1 + \left(
{\eta - s_{0}}  \right)}}{{1 - \left( {\eta - s_{0}}  \right)}}} \right\}
 =
\frac{{\alpha _{1}} }{{2}}{\cal F} \ln\frac{{2}}{{\delta s}},
\end{gather*}

\noindent whereof

\begin{gather}
\delta s =
2\left\{ {\frac{{1 + \left( {\eta -
s_{0}}  \right)}}{{1 - \left( {\eta - s_{0}}  \right)}}}
\right\}^{\frac{{\alpha _{2}} }{{\alpha _{1} }}\left( {\eta -
s_{0}}  \right)}\times\nonumber\\
\times\exp\left\{ { - 2\left( {1 + \frac{{\alpha
_{2}} }{{\alpha _{1} \eta} }} \right)} \right\}
\exp\left\{ { - \frac{{2}}{{\alpha _{1} {\cal F}}}}
\right\},
\label{eq37}
\end{gather}

\noindent where

\[
s = 1 + \delta s.
\]

\noindent Taking into account (\ref{eq33}), for conditions
(\ref{eq34}), (\ref{eq35}) we have

\[
\left\{ {\frac{{\partial \tilde {\varepsilon} _{1}} }{{\partial
\omega} }} \right\}_{\omega = \omega _{0}}  \approx
\frac{{1}}{{\omega _{0}} }\alpha _{1} \frac{{{\cal F}}}{{2\delta
s}}{\begin{array}{*{20}c}
 {,} \hfill & {\omega _{0} = kv_{1F}}  \hfill \\
\end{array}} .
\]

\noindent Further using formulas (\ref{eq31}), (\ref{eq36}),
(\ref{eq37}), we have arrived at the following relation for
$\gamma \left( {k} \right)$:

\begin{gather}
\frac{{\gamma \left( {k} \right)}}{{\omega _{0}} }
\approx 2\pi \frac{{\alpha _{2}} }{{\eta \alpha _{1}} }\left(
{\eta - s_{0}} \right)\times\nonumber\\
\times\left\{ {\frac{{1 + \left( {\eta - s_{0}}
\right)}}{{1 - \left( {\eta - s_{0}}  \right)}}}
\right\}^{\frac{{\alpha _{2}} }{{\alpha _{1}
}}\left( {\eta - s_{0}}  \right)}\times\nonumber\\
\times\exp\left\{ { - \frac{{2}}{{\alpha _{1} {\cal F}}}\left(
{1 + \alpha _{1} {\cal F} + \frac{{\alpha _{2} {\cal F}}}{{\eta}
}} \right)} \right\},
\label{eq38}
\end{gather}

\[
\left| {\eta - s_{0}}  \right| < 1, \quad \omega _{0} = kv_{1F} ,
\quad {\cal F} \ll 1.
\]

According to (\ref{eq4}), (\ref{eq24}) the condition ${\gamma
\left( {k} \right) > 0}$, or $0 < \left( {\eta - s_{0}}  \right) <
1$, corresponds to the oscillation damping, whereas the condition
$\gamma \left( {k} \right) < 0$, or $ - 1 < \left( {\eta - s_{0}}
\right) < 0$ corresponds to the oscillation increase, meaning
instability development. The latter inequality can be written (in
view of (\ref{eq32})) as $\eta < s_{0} < \eta + 1$, or taking into
account $\eta = \sqrt {{{\varepsilon _{1F}}  \mathord{\left/
{\vphantom {{\varepsilon _{1F}}  {\varepsilon _{2F}} }} \right.
\kern-\nulldelimiterspace} {\varepsilon _{2F}} }} $

\begin{equation}
\label{eq39} v_{1F} < u\cos\alpha < v_{1F} + v_{2F} ,
\end{equation}

\noindent where $\alpha $ is an angle between vectors $\mathbf{u}$
and $\mathbf{k}$.

It is evident from (\ref{eq38}) that when approaching to the right
boundary of interval (\ref{eq39}), that is if

\begin{equation}
\label{eq40} s_{0} \to \eta + 1{\begin{array}{*{20}c}
 {,} \hfill & {s = 1 + \delta s} \hfill \\
\end{array}}
\end{equation}

\noindent the increment $\gamma \left( {k} \right)$ is
unrestrictedly increasing. However, it is evident that for the
case (\ref{eq40}) we need to solve more correctly dispersion
equation (\ref{eq30}), which in this approximation and keeping in
mind $s \to 1$ has the form:

\[
1 + \alpha _{1} {\cal F} + \frac{{\alpha _{2}} }{{\eta} }{\cal F}
= - \frac{{1}}{{2}}\alpha _{1} {\cal F} \ln\frac{{\delta s}}{{2}}
- \frac{{\alpha _{2} }}{{2\eta} }{\cal F} \ln\frac{{\eta \delta
s}}{{2}}.
\]

\noindent The solution of this equation is given by the relation:

\begin{gather}
\delta s = 2\eta ^{ - \dfrac{{\alpha _{2}} }{{\eta
_{1} \alpha _{1} + \alpha
_{2}} }}\times\\
\times\exp\left\{ {  \frac{{-2\eta } }{({\eta \alpha
_{1} + \alpha _{2}})\cal F }\left[ { {1
+ \alpha _{1} \cal F + \frac{{\alpha _{2}
\cal F}}{{\eta} }} } \right]} \right\}.\nonumber
\label{eq41}
\end{gather}

Remembering that according to (\ref{eq33}) under conditions
(\ref{eq31}), (\ref{eq36}), (\ref{eq40}) the relation for $\left\{
{\dfrac{{\partial \tilde {\varepsilon} _{1} }}{{\partial \omega}
}} \right\}_{\omega = \omega _{0}}  $ is

\[
\left\{ {\frac{{\partial \tilde {\varepsilon} _{1}} }{{\partial
\omega} }} \right\}_{\omega = \omega _{0}}  \approx
\frac{{1}}{{\omega _{0} }}\frac{{\eta \alpha _{1} + \alpha _{2}}
}{{\eta} }\frac{{{\cal F}}}{{2\delta s}}
\]

\noindent and considering (\ref{eq31}), (\ref{eq36}),
(\ref{eq41}),
 we can obtain the following formula for
increment~$\gamma \left( {k} \right)$:

\begin{gather}
\frac{{\gamma \left( {k} \right)}}{{\omega _{0}} } \approx -
2\pi \frac{{\alpha _{2}} }{{\eta \alpha _{1} + \alpha _{2}} }\eta
^{ -
\dfrac{{\alpha _{2}} }{{\eta \alpha _{1} + \alpha _{2}} }}\times\nonumber\\
\times\exp\left\{ {  \frac{{-2\eta } }{({\eta \alpha
_{1} + \alpha _{2}})\cal F }\left[ { {1
+ \alpha _{1} \cal F + \frac{{\alpha _{2}
\cal F}}{{\eta} }} } \right]} \right\},\nonumber\\
\label{eq42}
\end{gather}

\[
 {\cal F} \ll 1, \quad \omega _{0} \approx kv_{1F} ,\quad
 s \approx  1,\quad
s_{0} \lesssim \eta + 1,
\]

\noindent where quantities $s_{0} ,\;\;\eta ,\;\;s$ are defined as
before
 by relations (\ref{eq32}). Taking into account that
$\left( {\eta \alpha _{1} /\eta \alpha _{1} + \alpha _{2}}
\right) < 1$, comparing relations (\ref{eq38}) and (42) one can
easily conclude that when ${\cal F}  \ll $ 1, the maximum value of
the increment is reached for the upper limit of the interval
(\ref{eq39}), i.e.,when

\begin{equation}
\label{eq43} \sqrt {E} \cos\alpha \lesssim \sqrt {\varepsilon
_{1F}}  + \sqrt {\varepsilon _{2F}}  ,
\end{equation}

\noindent where \textit{E}  is the kinetic energy per one nucleon
in the impacting drop.

\section{Temperature impact}

So far we have discussed the problem of propagation of weakly
increasing or damping zero-sound oscillations, assuming that the
colliding Fermi-liquid drops have zero temperature~(see
(\ref{eq14})). However, when two equilibrium Fermi-liquid drops
collide and the temperature of each one is non-zero, a scenario of
weak instability development can have significant peculiarities.

In order to solve this problem we need to modify relations
(\ref{eq30}), (\ref{eq31}), determining frequency and increments
(decrements) of zero-sound oscillations in conformity with the
case of the colliding equilibrium Fermi-liquid drops with the
non-zero temperature. In this case the equilibrium distribution
functions of the drops are as follows~(see (\ref{eq8}) -
(\ref{eq13})):

\[
f_{0}^{\left( {1} \right)} \left( {\mathbf{p}} \right) = \left\{
{\exp\left[ {{{\left( {\frac{{\mathbf{p}^{2}}}{{2m}} - \varepsilon
_{1F}}  \right)} \mathord{\left/ {\vphantom {{\left(
{\frac{{p^{2}}}{{2m}} - \varepsilon _{1F}}  \right)} {T_{1}} }}
\right. \kern-\nulldelimiterspace} {T_{1}} }} \right] + 1}
\right\}^{ - 1},
\]

\begin{gather}
f_{0}^{\left( {2} \right)} \left( {\mathbf{p} - m\mathbf{u}} \right)=\nonumber\\
= \left\{ {\exp\left[ {{{\left( {\frac{{\left( {\mathbf{p} -
m\mathbf{u}} \right)^{2}}}{{2m}} -
 \varepsilon _{2F}}
\right)} \mathord{\left/ {\vphantom {{\left( {\frac{{\left(
{\mathbf{p} - m\mathbf{u}} \right)^{2}}}{{2m}} - \varepsilon
_{2F}}  \right)} {T_{2}} }} \right. \kern-\nulldelimiterspace}
{T_{2}} }} \right] + 1} \right\}^{ - 1},
\label{eq44}
\end{gather}

\noindent where \textit{T}$_{1}$, \textit{T}$_{2}$ are
temperatures of the resting and impacting drops, respectively. The
given below relations should hold true for the drops

\begin{equation}
\label{eq45}
 \left( {T_{1} /\varepsilon _{1F}}  \right) \ll  1,
 \quad \left( {T_{2} /\varepsilon _{2F}}  \right) \ll  1.
\end{equation}

The relations determine possibility to apply the Fermi-liquid
description to systems of strong interacting fermions (in the
present case - nucleons). Taking into account formulas
(\ref{eq44}) and inequalities (45), one can make low temperature
expansion in expression (\ref{eq23}), that can result in the
following equation for determination of frequency $\omega _{0} =
skv_{1F} $ of the zero-sound oscillations in the main
approximation over parameters~$\left( {T_{1} /\varepsilon _{1F}}
\right)$, $\left( {T_{2} /\varepsilon _{2F}}  \right)$ (see
\cite{yS}):

\[
\varepsilon _{1} \left( {s} \right) \equiv \tilde {\varepsilon}
_{10} \left( {s} \right) + \tilde {\varepsilon} _{1T} \left( {s}
\right) = 0,
\]

\end{multicols}

\begin{equation*}
\tilde {\varepsilon} _{10} \left( {s} \right) = 1 + \alpha _{1}
{\cal F}\left\{ {1 - \frac{{s}}{{2}}\ln\left| {\frac{{s + 1}}{{s -
1}}} \right|} \right\}
+\alpha _{2} \frac{{1}}{{\eta} }{\cal F}
\left\{ {1 - \frac{{1}}{{2}}\left( {\eta s - s_{0}}
\right)\ln\left| {\frac{{\eta s - s_{0} + 1}}{{\eta s - s_{0} -
1}}}
\right|} \right\} = 0,
\end{equation*}

\begin{gather}
\tilde {\varepsilon} _{1T} \left( {s} \right) =\nonumber\\
=\frac{{-\alpha _{1} {\cal F}}}{{2}} P\int\limits_{0}^{\infty}
{\frac{{z dz}}{{e^{z} + 1}}} \frac{{2 + \left( {s^{2} - 1}
\right)}}{{\left( {s^{2} - 1} \right)^{2}\left( {\varepsilon _{1F}
/T_{1}}  \right)^{2} - z^{2}}}
-\frac{{\alpha _{2}
{\cal F}}}{{2\eta} }P\int\limits_{0}^{\infty} {\frac{{z dz}}{{e^{z}
+ 1}}} \, \frac{{2 + \left[ {\left( {\eta s - s_{0}} \right)^{2} -
1} \right]}}{{\left[ {\left( {\eta s - s_{0}}  \right)^{2} - 1}
\right]\left( {\varepsilon _{2F} /T_{2}}  \right)^{2} - z^{2}}},
\label{eq46}
\end{gather}

\begin{multicols}{2}

\noindent where \textit{P} means as before the symbol of the
principal value. In the main approximation the expression for
quantity $\tilde {\varepsilon} _{2} \left( {s} \right)$, defining
in accordance with (\ref{eq27}), (\ref{eq31}) a weak increase or
damping of the zero-sound oscillations, can be put as:


\begin{gather}
\tilde {\varepsilon} _{2} \left( {s} \right) =
\frac{{\pi} }{{2}}{\cal F}\left\{ \alpha _{1} sf_{0} \left(
{\frac{{\varepsilon _{1F}} }{{T_{1}} }\left( {s^{2} - 1} \right)}
\right) +\right.\nonumber\\
\left.+\frac{{\alpha _{2}} \left( {\eta s - s_{0}}
\right)}{{\eta} }
f_{0} \left( {\frac{{\varepsilon _{2F}} }{{T_{2}} }\left[
{\left( {\eta s - s_{0}}  \right)^{2} - 1} \right]} \right)
\right\},
\label{eq47}
\end{gather}

\noindent where

\begin{equation}
\label{eq48} f_{0} \left( {\varepsilon /T} \right) = \left\{
{e^{\varepsilon /T} + 1} \right\}^{ - 1}.
\end{equation}

It is easy to see that at $T_{1} \to  0, \, T_{2} \to 0$ quantity
$\tilde {\varepsilon} _{1T} \left( {s} \right)$ turns to zero.
Hence, equation (\ref{eq46}) transforms into dispersion equation
(\ref{eq30}). Expression (\ref{eq47}), with account of
(\ref{eq48}), transforms
 to formula (\ref{eq31}) for quantity $\tilde
{\varepsilon} _{2} \left( {s} \right)$, determining instabilities
of oscillations when the drops have zero temperature. As we can
see from (\ref{eq46}), (\ref{eq47}), the temperature effects can
greatly impact development of the instabilities associated with
propagation of the zero-sound oscillations in the system, when $s
\approx 1$, i.e. ${\cal F} \ll  1$. Keeping in mind this fact, let
us analyze the possibility of weak increase of the zero-sound
oscillations in the system when the resting Fermi-liquid drop is
found at zero temperature, \textit{T}$_{1}$=0. Here we will also
assume that the following relations (compare with с (\ref{eq34}),
(\ref{eq35}), (\ref{eq40}), (\ref{eq43}), (45)) are valid

\[
 {\cal F}  \ll  1,\quad s\gtrsim 1, \quad s_{0} \lesssim \eta + 1,\quad  \left(
{\varepsilon _{2F} /T_{2}}  \right) \gg 1,
\]

\begin{equation}
\label{eq49} \delta s\left( {\varepsilon _{2F} /T_{2}}  \right)
\ll 1, \quad \delta s = s - 1 \ll 1.
\end{equation}

Under these conditions dispersion equation (\ref{eq46}), meeting
these conditions, in the main approximation over $\delta s$ will
get the form:

\begin{gather}
\tilde {\varepsilon} _{1} \left( {s} \right)
\approx  1 + \alpha _{1} {\cal F} + \alpha _{2} {\cal F}\left( {1
- I_{0}}  \right) -\nonumber\\
 -\frac{{\alpha _{2} {\cal F}}}{{\eta
}}\ln\frac{{4\varepsilon _{2F}} }{{T_{2}} }
+ \frac{{\alpha _{1} {\cal F}}}{{2}}\ln\frac{{\delta s}}{{2}} =
0.
\label{eq50}
\end{gather}

In order to obtain this equation we used the asymptotic estimate
of the integral, when $|t|\ll 1$

\begin{equation}
\label{eq51} P\int\limits_{0}^{\infty}  {\frac{{zdz}}{{\left(
{e^{z} + 1} \right)\left( {t^{2} - z^{2}} \right)}}}
\mathrel{\mathop{\kern0pt\longrightarrow}\limits_{\left| {t}
\right| \ll 1}} I_{0} + \frac{{1}}{{2}}\ln\left| {t} \right| -
I_{1} t^{2},
\end{equation}

\begin{gather*}
I_{0} = \frac{{1}}{{4}} - \int\limits_{1}^{\infty}
{\frac{{dz}}{{z\left( {e^{z} + 1} \right)}}} +\\
+\frac{{1}}{{2}}\int\limits_{0}^{1} {\frac{{dz}}{{z}}\left\{
{\tanh\frac{{z}}{{2}} - \frac{{z}}{{2}}} \right\}
\approx 0,07} ,
\end{gather*}

\[
I_{1} = \int\limits_{1}^{\infty}  {\frac{{dz}}{{z^{3}\left( {e^{z}
+ 1} \right)}} -}  \frac{{1}}{{2}}\int\limits_{0}^{1}
{\frac{{dz}}{{z^{3}}}\left\{ {\tanh\frac{{z}}{{2}} - \frac{{z}}{{2}}}
\right\} \approx 0,11} .
\]

\noindent The integrals having exactly the same form describe the
temperature impact on dispersion equation (\ref{eq46}).
 The solution of equation (\ref{eq50}) is:

\begin{equation}
\label{eq52} \delta s = 2\left( {4\dfrac{{\varepsilon _{2F}}
}{{T_{2}} }} \right)^{\dfrac{{\alpha _{2}} }{{\eta \alpha _{1}} }}
\times\exp\left\{ { - \frac{{2Q_{1}}}{{\alpha _{1} {\cal F}}}}
\right\},
\end{equation}

\begin{equation*}
Q_{1}\equiv{1 + \alpha _{1} {\cal F} + \frac{{\alpha _{2} {\cal
F}}}{{\eta} }\left( {1 - I_{0}}  \right)}.
\end{equation*}

\noindent In accordance with (\ref{eq49}) this solution must
satisfy relation $\delta s\left( {\varepsilon _{2F} /T_{2}}
\right) \ll 1$, whence, remembering that $\left( {\varepsilon
_{2F} /T_{2}}  \right)~\gg~1$, it is easy to determine a
``temperature'' range, when solution (\ref{eq52}) exists for
$\delta s$:

\begin{equation}
\label{eq53} 1 \ll \frac{{\varepsilon _{2F}} }{{T_{2}} } \ll
\exp{\left\{ {\frac{{2\eta Q_{1} }}{{\left( {\eta \alpha _{1} +
\alpha _{2}}  \right){\cal F}}}} \right\}}.
\end{equation}

Noting further that under (\ref{eq50}) the formula holds

\[
\left\{ {\frac{{\partial \tilde {\varepsilon} _{1}} }{{\partial
\omega} }} \right\}_{\omega = \omega _{0}}  \approx
\frac{{1}}{{\omega _{0} }}\frac{{\alpha _{1} {\cal F}}}{{2\delta
s}}, \quad \omega _{0} = kv_{1F}
\]

\noindent and using the fact that when \textit{T}$_{1}   \to $ 0
and $s\gtrsim1$

\begin{equation}
\label{eq54} \tilde {\varepsilon} _{2} = - \frac{{\pi}
}{{4}}\frac{{\alpha _{2}} }{{\eta }}{\cal F},
\end{equation}

\noindent we can arrive at the following expression for increment
$\gamma _{k} \left( {s} \right)$, considering~(\ref{eq31}),

\begin{gather}
\frac{{\gamma _{k}} }{{\omega _{0}} } = - \pi
\frac{{\alpha _{2}} }{{\eta \alpha _{1}} }\left(
{4\frac{{\varepsilon _{2F}} }{{T_{2}} }} \right)^{\dfrac{{\alpha
_{2}} }{{\eta \alpha _{1}} }}\times\nonumber\\
\times\exp\left\{ { - \frac{{2}}{{\alpha _{1} {\cal F}}}\left[
{1 + \alpha _{1} {\cal F}
 + \frac{{\alpha _{2}
{\cal F}}}{{\eta} }\left( {1 - I_{0}}  \right)} \right]} \right\},\nonumber\\
\label{eq55}
\end{gather}

\noindent which is true at the conditions (\ref{eq49}),
(\ref{eq53}).
 In order to obtain formula
(\ref{eq54}), we have used expression (\ref{eq47}), and also have
taken into account that (see~(\ref{eq48}))

\[
 f_{0} \left( {\frac{{\varepsilon _{1F}} }{{T_{1}} }\left( {s^{2} - 1}
\right)}
\right)\mathrel{\mathop{\kern0pt\longrightarrow}\limits_{T_{1} \to
0}} \theta \left( {1 - s^{2}} \right) = 0,\, s\gtrsim 1,
\]

\begin{gather*}
f_{0} \left[ {\frac{{\varepsilon _{2F}} }{{T_{2}} }\left( {\eta s
- s_{0}} \right)^{2} - 1} \right]\approx\\
 \approx f_{0} \left(
{2\frac{{\varepsilon _{2F}} }{{T_{2}} }\eta \delta s} \right)
\approx f_{0} \left( {0} \right) = \frac{{1}}{{2}},
\end{gather*}

\[
 s_{0}
\lesssim \eta + 1.
\]

Now let us consider a case when the impacting Fermi-liquid drop
has zero temperature, \textit{T}$_{2}$=0, and the resting drop has
temperature \textit{T}$_{1.}$ We assume that conditions (compare
with (\ref{eq49})) are valid:

\[{\cal F}  \ll  1,\quad s\gtrsim 1, \quad s_{0} \lesssim \eta
+ 1,\quad \left( {\varepsilon _{1F} /T_{1}}  \right) \gg 1,
\]

\begin{equation}
\label{eq56} \delta s\left( {\varepsilon _{1F} /T_{1}}  \right)
\ll 1,
 \quad \delta s = s - 1 \ll 1.
\end{equation}

\noindent In this case the dispersion equation in the main
approximation over $\delta s$ considering asymptotic estimate
(\ref{eq51}) has the form:

\begin{gather}
\tilde {\varepsilon} _{1} \left( {s} \right)
\approx 1 + \alpha _{1} {\cal F}\left( {1 - I_{0}}  \right) +
\frac{{\alpha _{2} {\cal F}}}{{\eta} } -\nonumber\\
- \frac{{\alpha _{1} {\cal
F}}}{{2}}\ln\frac{{4\varepsilon _{1F}} }{{T_{1}} }
+ \frac{{\alpha _{2} {\cal F}}}{{2\eta} }\ln\frac{{\eta\delta
s}}{{2}} = 0.
\label{eq57}
\end{gather}

\noindent A solution of this equation is given by

\begin{equation}
\delta s = \frac{{2}}{{\eta} }\left( {4\frac{{\varepsilon _{1F}}
}{{T_{1} }}} \right)^{\dfrac{{\eta \alpha _{1}} }{{\alpha _{2}} }}
\times\exp\left\{ { - \frac{{2\eta Q_{2}} }{{\alpha _{2} {\cal
F}}}} \right\}, \label{eq58}
\end{equation}

\begin{equation*}
Q_{2}\equiv {1 + \alpha _{1} {\cal F}\left( {1 - I_{0}} \right) +
\frac{{\alpha _{2} {\cal F}}}{{\eta} }},
\end{equation*}

\noindent and here, we should remember that in accordance with
(\ref{eq56}) the relation $\delta s\left( {\varepsilon _{1F}
/T_{1}}  \right) \ll 1$ must be valid, and a ''temperature''
condition of existence of such solution is determined by the
inequality (compare with ~(\ref{eq53})):

\begin{equation}
\label{eq59} 1 \ll \frac{{\varepsilon _{1F}} }{{T_{1}} } \ll
\exp{\left\{ {\frac{{2\eta Q_{2}}}{{\left( {\eta \alpha _{1} +
\alpha _{2}}  \right){\cal F}}}} \right\}}.
\end{equation}

Noting further that for  $s\gtrsim 1, \, s_{0} \lesssim \eta + 1 $
and $ T_{2} = 0$ the following formulas are valid

\[
f_{0} \left( {\frac{{\varepsilon _{1F}} }{{T_{1}} }\left( {s^{2} -
1} \right)} \right) \approx f_{0} \left( {2\frac{{\varepsilon
_{1F}} }{{T_{1} }}\delta s} \right) \approx f_{0} \left( {0}
\right) = \frac{{1}}{{2}},
\]

\[
f_{0} \left( {\frac{{\varepsilon _{2F}} }{{T_{2}} }\left[ {\left(
{\eta s - s_{0}}  \right)^{2} - 1} \right]}
\right)\mathrel{\mathop{\kern0pt\longrightarrow}\limits_{T_{2} \to
0}} \theta \left( {1 - \left( {\eta s - s_{0}}  \right)^{2}}
\right) ,
\]

\[
f_{0} \left( {\frac{{\varepsilon _{2F}} }{{T_{2}} }\left[ {\left(
{\eta s - s_{0}}  \right)^{2} - 1} \right]}
\right)\mathrel{\mathop{\kern0pt\longrightarrow}\limits_{T_{2} \to
0}}  1,
\]

\noindent the expression for $\tilde {\varepsilon} _{2} $,
determining in accordance with (\ref{eq31}) the damping or
increase of the zero-sound oscillations, can be put in a form

\begin{equation}
\label{eq60} \tilde {\varepsilon} _{2} = \frac{{\pi} }{{4\eta}
}{\cal F}\left\{ {\eta \alpha _{1} - 2\alpha _{2}}  \right\}.
\end{equation}

Bearing in mind that quantity $\left\{ {\partial \tilde
{\varepsilon} _{1} /\partial \omega}  \right\}_{\omega = \omega
_{0}}  $ in compliance with (\ref{eq57}) is given by the formula

\begin{equation}
\label{eq61} \left\{ {\frac{{\partial \tilde {\varepsilon} _{1}}
}{{\partial \omega} }} \right\}_{\omega = \omega _{0}}  \approx
\frac{{1}}{{\omega _{0} }}\frac{{\alpha _{2} {\cal F}}}{{2\eta}
}\frac{{1}}{{\delta s}},
\end{equation}

\noindent where $\delta s$ is determined by (\ref{eq58}), and
substituting (\ref{eq60}), (\ref{eq61}) into (\ref{eq31}), we have

\begin{gather}
\frac{{\gamma _{k}} }{{\omega _{0}} } \approx \pi
\frac{{\eta \alpha _{1} - 2\alpha _{2}} }{{\eta \alpha _{2}}
}\left( {4\frac{{\varepsilon _{1F}
}}{{T_{1}} }} \right)^{\frac{{\eta \alpha _{1}} }{{\alpha _{2}} }}\times\nonumber\\
\times\exp\left\{ { - \frac{{2}}{{\alpha _{2} {\cal F}}}\left[
{1 + \alpha _{1} {\cal F}\left( {1 - I_{0}}
\right) + \frac{{\alpha _{2} {\cal F}}}{{\eta} }} \right]} \right\},
\label{eq62}
\end{gather}

\[
\omega _{0} = kv_{1F}.\nonumber
\]

As is easy to see, the formula (\ref{eq62}) in contrast to
(\ref{eq55}),
 which is correct when
$T_{1}=0, T_{2} \ne 0$, determines either a coefficient of damping
(when $\eta \alpha _{1} > 2\alpha _{2} $), or increase (when $\eta
\alpha _{1} < 2\alpha _{2} $) of the zero-sound oscillations. Such
non-invariance of (\ref{eq55}), (\ref{eq62}) relative to
interchanging of the impacting and resting drops with a
corresponding change of their characteristics  $\varepsilon _{1F}
\leftrightarrow \varepsilon _{2F}$, $ T_{1} \leftrightarrow T_{2}
$ (in other words the absence of the Galilean invariance in the
system of the collided drops) is not in any sense paradoxical.
This is because within the developed Fermi-liquid description
model the interaction of quasiparticles is not invariant with
respect to the Galilean transformations (see (\ref{eq2}),
(\ref{eq3})).

Therefore, the relation $\eta \alpha _{1} < 2\alpha _{2} $ that can be
written as

\begin{equation}
\label{eq63} \alpha _{1} \sqrt{\varepsilon _{1F}} < 2\alpha _{2}
\sqrt{\varepsilon _{2F}} ,
\end{equation}

\noindent along with (\ref{eq56}), (\ref{eq59}) determines
existence conditions of weakly increasing oscillations with
increment $\gamma _{k} $ in the system

\begin{gather}
\frac{{\gamma _{k}} }{{\omega _{0}} } \approx \pi
\frac{{\eta \alpha _{1} - 2\alpha _{2}} }{{\eta \alpha _{2}}
}\left( {4\frac{{\varepsilon _{1F}
}}{{T_{1}} }} \right)^{\dfrac{{\eta \alpha _{1}} }{{\alpha _{2}} }}\times\nonumber\\
\times\exp\left\{ { - \frac{{2\eta} }{{\alpha _{2} {\cal
F}}}\left[ {1 + \alpha _{1} {\cal F}\left( {1 -
I_{0}}  \right) + \frac{{\alpha _{2} {\cal F}}}{{\eta} }} \right]} \right\}.\nonumber\\
\label{eq64}
\end{gather}

In order to finish studying the temperature impact on the
instability development in the system of two collided Fermi-liquid
drops, we want to obtain an expression for the increment when both
the resting and impacting drops have comparable ($T_{1} \sim T_{2}
$) nonzero temperatures, moreover,

\begin{gather*}
{\cal F} \ll  1, \, s \gtrsim 1,\,  s_{0} \lesssim \eta + 1,\,\\
 \left(
{\varepsilon _{1F} /T_{1}}  \right) \gg 1,\,  \left( {\varepsilon
_{2F} /T_{2}}  \right) \gg 1,
\end{gather*}

\begin{gather}
\delta s\left( {\varepsilon _{1F} /T_{1}}  \right)
\ll 1, \quad \delta s\left( {\varepsilon _{2F} /T_{2}}  \right)
\ll 1,\nonumber\\
 \quad \delta s = s - 1 \ll 1.
\label{eq65}
\end{gather}

Dispersion equation (\ref{eq46}) in the main approximation, with
taking into account conditions (\ref{eq65}), has the form:

\begin{equation}
\label{eq66} \tilde {\varepsilon} _{1} = B\delta s^{2} - A\delta s
+ 1 - A = 0,
\end{equation}

\noindent where

\[
A \equiv \frac{{{\cal F}}}{{2}}\left\{ {\alpha _{1}
\ln\frac{{\varepsilon _{1F} }}{{T_{1}} } + \alpha _{2}
\ln\frac{{\varepsilon _{2F}} }{{T_{2}} }} \right\},
\]

\begin{equation}
\label{eq67} B \equiv 4{\cal F}I_{1} \left\{ {\alpha _{1} \left(
{\frac{{\varepsilon _{1F} }}{{T_{1}} }} \right)^{2} + \eta \alpha
_{2} \left( {\frac{{\varepsilon _{2F}} }{{T_{2}} }} \right)^{2}}
\right\}
\end{equation}

\noindent and coefficient $I_{1} $ is determined by (\ref{eq51}).

The solution of equation (\ref{eq66}), since $(A/B) \ll 1$, will
be unambiguous and positive when~$A \ge 1$:

\begin{equation}
\label{eq68} \delta s = \frac{{A}}{{B}} + \sqrt
{\frac{{A^{2}}}{{4B^{2}}} + \frac{{A - 1}}{{B}}} , \quad A  \ge
1.
\end{equation}
Focusing on the explicit form of coefficients \textit{A} and
\textit{B,}  it is easy to prove that the conditions

\[
\delta s\left( {\varepsilon _{1F} /T_{1}}  \right) \ll 1, \quad
\delta s\left( {\varepsilon _{2F} /T_{2}}  \right) \ll 1
\]

\noindent are automatically fulfilled.

Proceeding from (\ref{eq31}), (\ref{eq33}) and noting that in
accordance with (\ref{eq47}), (\ref{eq65}), (\ref{eq66}) the
following formulas hold

\[
\tilde {\varepsilon} _{2} \approx \frac{{\pi} }{{4\eta} }{\cal
F}\left\{ {\eta \alpha _{1} - \alpha _{2}}  \right\},
\]
\[
\left\{ {\frac{{\partial \tilde {\varepsilon} _{1}} }{{\partial
\omega} }} \right\}_{\omega = \omega _{0}}  \approx
\frac{{1}}{{\omega _{0}} }\sqrt {A^{2} + 4B\left( {A - 1} \right)}
,
\]

\noindent we will attain the following expression for coefficient
$\gamma _{k} $:

\begin{equation}
\label{eq69} \frac{{\gamma _{k}} }{{\omega _{0}} } \approx
\frac{{\pi} }{{4\eta }}{\cal F}\frac{{\eta \alpha _{1} - \alpha
_{2}} }{{\sqrt {A^{2} + 4B\left( {A - 1} \right)}} } \ll 1, \quad
\omega _{0} = kv_{1F} .
\end{equation}

It is clear that when
$\eta \alpha _{1} < \alpha_{2} $, or

\begin{equation}
\label{eq70} \alpha _{1} \sqrt{\varepsilon _{1F}} < \alpha _{2}
\sqrt{\varepsilon _{2F}} ,
\end{equation}

\noindent quantity $\gamma _{k} $ represents an instability
increment of the zero-sound oscillations.

Expression  (\ref{eq69}) takes  a simple form

\begin{equation}
\label{eq71} \frac{{\gamma _{k}} }{{\omega _{0}} } \approx
\frac{{\pi} }{{4\eta }}{\cal F}\frac{{\eta \alpha _{1} - \alpha
_{2}} }{{\alpha _{1} \ln\left( {\varepsilon _{1F} /T_{1}}  \right)
+ \alpha _{2} \ln\left( {\varepsilon _{2F} /T_{2}}  \right)}}
\end{equation}

\noindent or

\[
\frac{{\gamma _{k}} }{{\omega _{0}} } \approx \frac{{\pi}
}{{4\eta} }{\cal F}\left( {\eta \alpha _{1} - \alpha _{2}}
\right), \quad \omega _{0} = kv_{1F}
\]

\noindent when $A \approx  1$, that is (see (\ref{eq67}))

\begin{equation}
\label{eq72} \alpha _{1} \ln\left( {\varepsilon _{1F} /T_{1}}
\right) + \alpha _{2} \ln\left( {\varepsilon _{2F} /T_{2}}
\right) \approx \frac{{2}}{{{\cal F}}}.
\end{equation}

For a more general case, i.e. when $A \gg 1$ and $A \ll B$ (see
(\ref{eq67})), formula (\ref{eq69}) also gets simpler, but it is
still more complicated compared to~(\ref{eq71}):

\end{multicols}
\begin{equation}
\label{eq73} \frac{{\gamma _{k}} }{{\omega _{0}} } \approx
\frac{{\pi\left( {\eta \alpha _{1} - \alpha
_{2}} \right)} }{{32\eta I_{1} }}\left\{ {\frac{{{\cal F}}}{{2}}\left[ {\alpha _{1}
\ln\left( {\varepsilon _{1F} /T_{1}} \right) + \alpha _{2}
\ln\left( {\varepsilon _{2F} /T_{2}}  \right)} \right] - 1}
\right\}^{ - 1/2} \left\{ {\alpha _{1} \left( {\frac{{\varepsilon
_{1F}} }{{T_{1}} }} \right)^{2} + \eta \alpha _{2} \left(
{\frac{{\varepsilon _{2F}} }{{T_{2} }}} \right)^{2}} \right\}^{ -
1}.
\end{equation}
\begin{multicols}{2}

Expressions for increments (\ref{eq71}), (\ref{eq73}) look even
simpler, if we consider collision of identical nuclei having the
same temperatures (see below formulas (\ref{eq75}), (\ref{eq76})).
Comparison of expressions (\ref{eq38}), (\ref{eq42}) with
(\ref{eq55}),
 (\ref{eq62}), (\ref{eq71}), (\ref{eq73}), with taking
into account relations (\ref{eq63}), (\ref{eq70}) allows to come
to a conclusion that for the colliding Fermi-liquid drops with
nonzero temperatures the instability increments are higher
compared to the ones for the colliding drops with zero
temperatures. In particular, for the collision of the Fermi-liquid
drops having comparable temperatures \textit{T}$_{1}\sim
$\textit{T}$_{2}$ ($\left( {T_{1} /\varepsilon _{1F}}  \right) \ll
1$, $\left( {T_{2} /\varepsilon _{2F}}  \right) \ll 1$),
 when
condition (\ref{eq72}) is fulfilled, the increase increment of the
zero-sound oscillations has
 a power like character of dependence
over the Landau amplitude ${\cal F} \ll 1$, whereas for other
considered here cases (except the one leading to (\ref{eq73})),
the degree of the increment smallness
 is given by
exponential multipliers of the type $\exp\left( { - \lambda /{\cal
F}} \right)$, ${\cal F} \ll 1$, where $\lambda $ is a certain
constant. If we recall that in our consideration collision of two
fast (but not relativistic) nuclei is simulated by collision of
the two Fermi-liquid drops, then the provided comment means that
the instabilities, related to increase of the zero-sound
oscillations in the system of the two collided excited nuclei
(i.e. nuclei with nonzero temperatures), must develop much more
intensively compared to the instabilities in the system of the
collided unexcited nuclei.

\section{Discussion of results}
Up to now we have studied the case of small Landau amplitudes
${\cal F} \ll 1$. As it was proved in this case the dispersion
equations of the zero-sound oscillations allow for an analytical
solution in the perturbation theory over small ${\cal F}$.
Analytical solutions of dispersion equations can be obtained also
for the case of large Landau amplitudes, ${\cal F} \gg 1$.
However, if we consider these solutions from the viewpoint of the
problem about instability development in the system of collided
non-relativistic nuclei, it would be easy to prove that a solution
of the dispersion equations in the perturbation theory over a
small parameter $1/{\cal F}$ will lead us beyond the
non-relativistic approximation, which was one of the main
assumptions in the present work.

Naturally, it is easy to prove that the solutions of dispersion
equation (\ref{eq30}) are within the range of large $s, s \gg 1$
(note at once that when $s \gg 1$, it is possible to neglect
temperature dependent components in equation (\ref{eq46})). But in
accordance with (\ref{eq31}), (\ref{eq33}), (\ref{eq47}) for
existence of  instabilities the following condition must be
satisfied $s_{0} > \eta s$, or

\begin{equation}
\label{eq74} \sqrt {\varepsilon /\varepsilon _{F}}  \cos\alpha > s
\gg 1,
\end{equation}

\noindent where $\varepsilon $ is kinetic energy per one nucleon
in the impacting drop. In inequality (\ref{eq74}) we took into
account that for heavy nuclei ${\varepsilon _{1F}  \approx
\varepsilon _{2F}  = \varepsilon _{F}}$, meaning, $\eta \approx
1$. Remembering that for the heavy nuclei $\varepsilon _{F}
\approx $ 36 MeV, and the non-relativistic approximation requires
the nucleon kinetic energy to be small compared to the rest energy
$mc^{2}\sim $ 1 GeV, ($\varepsilon /mc^{2} \ll 1$), it is easy to
see that relation (\ref{eq74}) can not be satisfied under the
non-relativistic approximation. Therefore, in this work we do not
consider the case of large Landau amplitudes.

In this article we have restricted ourselves by consideration of
positive Landau amplitudes. When Landau amplitudes are negative,
the dispersion equation of the zero-sound oscillations can be
solved only by numerical methods. This is because in this case a
real and imaginary parts of the frequency are comparable with each
other and the analytical methods, based on the perturbation
theory, become inapplicable. The numerical solution of the
dispersion equations for the negative Landau amplitudes is not
included into the present work, which has as its object to
demonstrate a principal possibility of instability development in
the system of the colliding heavy nuclei, induced by propagation
and increase of the zero-sound oscillations.

Formulas (\ref{eq71}), (\ref{eq73}) could be used as a proof of
possible development of just such instabilities in the system of
the heavy colliding nuclei. When two identical nuclei collide
(${\varepsilon _{1F} = \varepsilon _{2F} = \varepsilon _{F}}$ ,$
\eta = 1$, $T_{1} \approx T_{2} = T$), formula (\ref{eq71}) takes
the form:

\begin{equation}
\label{eq75} \frac{{\gamma _{k}} }{{\omega _{0}} } \approx
\frac{{\pi }}{{2}}\frac{{\alpha _{1} - \alpha _{2}^{}} }{{\alpha
_{1} + \alpha _{2} }}\frac{{1}}{{\ln\left( {\varepsilon _{F} /T}
\right)}}, \quad \omega _{0} = kv_{F} .
\end{equation}

In this case the expression (\ref{eq73}) is significantly
simplified too:

\begin{gather}
\frac{{\gamma _{k}} }{{\omega _{0}} } \approx
\frac{{\pi} }{{32I_{1} }}\left(\frac{{\alpha _{1} - \alpha _{2}}
}{{\alpha _{1}^{} + \alpha _{2}
}}\right)\left( {\varepsilon _{F} /T} \right)^{-2}\times\nonumber\\
\times\left\{ {\left( {1/2} \right){\cal F}\left(
{\alpha _{1} + \alpha _{2}}  \right)\ln\left( {\varepsilon _{F}
/T} \right) - 1}
\right\}^{-1/2},\nonumber\\
\label{eq76}
\end{gather}

As it is we known temperature \textit{T} can be related to the
excitation energy of nucleus \textit{U} by an approximate
 formula (in this connection see, for
example,~\cite{jB,mP,pB,wH})

\begin{equation}
\label{eq77} T \approx \sqrt {aU} ,
\end{equation}

\noindent where $a$ is a certain constant, which should be
determined from experimental data (see in particular
\cite{jB,mP,pB}). For example, in accordance with \cite{jB} for a
nucleus with mass number $A = 115,  a \approx 1/8  \,$MeV;$ A =
181, a \approx 1/10 \, $MeV. Note that within the framework of the
Fermi-gas model of nucleus quantity $a$ can be found from
formula~\cite{wH}

\[
a = \frac{{4\varepsilon _{F}} }{{A\pi ^{2}}}.
\]

\noindent Following this formula for $\varepsilon _{F}  \approx $
36 MeV and $ A = 115, A = 181$, we have $a  \approx  0,124 \,$MeV
and $a
 \approx  0,0795\,$ MeV, respectively.

If we consider that the instabilities, associated with the
increase of amplitudes of the zero-sound oscillations, are
responsible for fragmentation, then specific time of fragmentation
$\tau _{f} $ must be of the order of the value reciprocal to
increment~$\gamma _{k} $

\begin{equation}
\label{eq78} \tau _{f} \sim 1/\gamma _{k}.
\end{equation}

\noindent Thus, if we could experimentally mearsure  the
fragmentation time of nuclei $\tau _{f} $, then the specific
dependence of the time from excitation energy \textit{U} would be
governed by the formula

\begin{equation}
\label{eq79} \tau _{f} \sim C\ln\left( {\frac{{\varepsilon _{F}}
}{{\sqrt {aU}} }} \right), \quad C = \omega _{0} \frac{{2}}{{\pi}
}\frac{{\left( {\alpha _{1} + \alpha _{2}} \right)}}{{\left(
{\alpha _{1} - \alpha _{2}}  \right)}}
\end{equation}

\noindent when relation (\ref{eq72}) holds, and by the formula

\begin{equation}
\label{eq80} \tau _{f} \sim C\left( {\frac{{\varepsilon _{F}}
}{{\sqrt {aU}} }} \right)^{2}\left\{ {{\cal F}\left( {\alpha _{1}
+ \alpha _{2}}  \right)\ln\left( {\frac{{\varepsilon _{F}}
}{{\sqrt {aU}} }} \right) - 2} \right\},
\end{equation}

\noindent when condition $A \gg $ 1, leading to (\ref{eq73}) is
valid. According to (\ref{eq75})-(\ref{eq78}) this would show that
the proposed mechanism of the instability development in the
system of the colliding heavy nuclei is working. Besides, if such
instability development mechanism is realized, it can result in
nucleus fragmentation and consequently, according to (\ref{eq72}),
(\ref{eq79}), (\ref{eq80}) we could estimate both values of
frequencies of oscillations of developed instabilities and
values of the Landau amplitudes for the nuclear matter formed as a
result of the collision. It is should be noted that the Landau
amplitudes are parameters of the Fermi-liquid approach of the
present work and, generally speaking, they don't necessarily
coincide with the Landau amplitudes for certain heavy nuclei,
values of which are contained in some published papers (see, for
example \cite{sB1,sB2}).

However, if we make an assumption that values of the Landau
amplitudes in the nuclear matter, formed as a result of the
collision of two fast heavy nuclei, are close to the corresponding
values of amplitudes of certain nuclei, then we can estimate
values of increments for specific cases and provide suggestions
about range of use for the results of the present work. Work
\cite{sB2} gives grounds to make a conclusion that the value of
Landau amplitude $F_{0}^{\left( {0} \right)} $, determining
oscillations of the nucleus density (see (\ref{eq6})), is within
the range of -0,25 and 0,5. These values of the Landau amplitudes
from the positive part of this range, when $\varepsilon _{1F}
\approx \varepsilon _{2F} = 36 \,MeV$, are fully within the frame
of the assumptions made in present work.

The values of amplitudes $F_{0}^{\left( {s} \right)}
,F_{0}^{\left( {i} \right)} $ (see (\ref{eq6})), determining
oscillations of the spin and isospin densities in the nucleus (see
\cite{sB1,sB2}) are also in compliance with the assumptions of our
work:

 $F_{0}^{\left( {s} \right)}  \approx  0,27  \div  0,35, F_{0}^{\left( {i}
\right)}  \approx  0,59  \div  0,72$.

As for the negative values of amplitude $F_{0} $ and values of
amplitude $F_{0}^{\left( {si} \right)} $ (see \cite{sB1,sB2})

 $F_{0}^{\left( {si} \right)}  \approx  1,26  \div  1,36$,

\noindent determining zero-sound oscillations of the spin-isospin
density, these cases as we have already said require numerical
calculations. This is the reason why they are not included into
this work.

Finally, let us study the question about peculiarities in behavior
of a non-equilibrium distribution function of two colliding
nucleon drops in the momentum space. This will allow to determine
a direction of nucleus fragment release as a result of instability
development in such a system with respect to a direction of
velocity of the moving drop.

It follows from (\ref{eq17}), that deviation of distribution
function $ \tilde {\delta f}\left(
{\mathbf{p},{\mathbf{k}},\omega} \right)$ from the equilibrium
distribution function is determined by the equation

\end{multicols}
\begin{gather}
 \tilde {\delta f}\left( {\mathbf{p},\mathbf{k},\omega}  \right) = - F\left\{ {\frac{{\alpha
_{1}} }{{\omega - \mathbf{k}\mathbf{v} +
io}}\mathbf{k}\frac{{\partial f_{0}^{\left( {1} \right)} \left(
{\mathbf{p}} \right)}}{{\partial \mathbf{p}}} + \frac{{\alpha
_{2}} }{{\omega - \mathbf{k}\mathbf{v} +
io}}\mathbf{k}\frac{{\partial f_{0}^{\left( {2} \right)} \left(
{\mathbf{p} - m\mathbf{u}} \right)}}{{\partial \mathbf{p}}}}
\right\}\times\nonumber\\
\times
\int {d\tau ^{'}}  \tilde {\delta f}\left(
{\mathbf{p}^{'},\mathbf{k},\omega}  \right) + \delta A\left(
{\mathbf{p},\mathbf{k}} \right)\delta \left( {\omega -
\mathbf{v}\mathbf{k}} \right),
\label{eq81}
\end{gather}
\begin{multicols}{2}

\noindent where derivatives $\dfrac{{\partial f_{0}^{\left( {1}
\right)} \left( {\mathbf{p}} \right)}}{{\partial \mathbf{p}}}$,
 $\dfrac{{\partial f_{0}^{\left( {2} \right)} \left(
{\mathbf{p} - m\mathbf{u}} \right)}}{{\partial \mathbf{p}}}$ are
given by (\ref{eq16}). According to (\ref{eq4}), (\ref{eq81}) the
non-equilibrium distribution function can have maximum, given the
following conditions are met

\begin{gather*}
\left. {\left( {\omega - \mathbf{k}\mathbf{v}} \right)}
\right|_{\mathbf{v} =
\mathbf{v}_{1F}}  = 0,\\
\left. {\left( {\omega - \mathbf{k}\mathbf{u} - \mathbf{k}\left(
{\mathbf{v} - \mathbf{u}} \right)} \right)} \right|_{\left|
{\mathbf{v} - \mathbf{u}} \right| = v_{2F}}  = 0,\,\\
 \mathbf{v} =
\mathbf{p}/m.
\end{gather*}

\noindent Since $\omega \approx \omega _{0} = skv_{1F} $, these
conditions can be written as

\begin{equation}
\label{eq82} s - \cos\theta = 0, \quad \eta s - s_{0} - \cos\beta
= 0,
\end{equation}

\noindent where $\theta $ is an angle between directions of
vectors $\mathbf{k}$ and $\mathbf{v}$, $\beta $ is an angle
between directions of vector $\mathbf{k}$ and vector $\mathbf{v} -
\mathbf{u}$. The first of these conditions is not fulfilled
because $s > 1$. As we have previously mentioned a damping or
increase of the zero-sound oscillations can take place only when
$\left| {\eta s - s_{0}}  \right| < 1$. This means that the second
condition in (\ref{eq82}) can be fulfilled. The maximum value of
increment $\gamma _{k} $ is reached when $\eta s - s_{0} \sim - 1$
(see (\ref{eq40})). Under condition $\varepsilon _{1F} \approx
\varepsilon _{2F} = \varepsilon _{F}  \left( {\eta = 1} \right)$
 and $s \gtrsim 1$ it corresponds to the condition

\begin{equation}
\label{eq83} s_{0} = \frac{{u}}{{v_{F}} }\cos\alpha \approx 2
\end{equation}

\noindent ($\alpha $ is as before an angle between directions of
vectors $\mathbf{u}$ and $\mathbf{k}$) and, hence

\begin{equation}
\label{eq84} u \ge 2v_{F}.
\end{equation}

Relation $\eta s - s_{0} \sim - 1$ is accordance with (\ref{eq82})
means that vectors $\mathbf{k}$ and $\mathbf{v} - \mathbf{u}$ are
anti-parallel

\[
\mathbf{v} - \mathbf{u} = - \left| {\mathbf{v} - \mathbf{u}}
\right|\frac{{\mathbf{k}}}{{k}},
\]

\noindent or, bearing in mind that $\left| {\mathbf{v} -
\mathbf{u}} \right| = v_{F} $,

\[
\mathbf{v} - \mathbf{u} = - v_{F} \frac{{\mathbf{k}}}{{k}}.
\]

Multiplying scalarwise this equation by vector $\mathbf{v} -
\mathbf{u}$, with taking into account (\ref{eq83}), we can obtain
 the following relation for cosine of angle $\theta _{0}
$ between directions of vectors $\mathbf{v}$ and $\mathbf{u}$

\[
\cos\theta _{0} = \frac{{u^{2} - 2v_{F}^{2}} }{{vu}},
\]

Hence, remembering that the module of vector $v$ is within the
region of values $u - v_{F} $ and $u + v_{F} $, and also that by
(\ref{eq84})

\[
\frac{{u^{2} - 2v_{F}^{2}} }{{u\left( {u + v_{F}}  \right)}} < 1,
\quad \frac{{u^{2} - 2v_{F}^{2}} }{{u\left( {u - v_{F}}  \right)}}
\ge 1,
\]

\noindent we have

\[
\frac{{u^{2} - 2v_{F}^{2}} }{{u\left( {u + v_{F}}  \right)}} \le
\cos\theta _{0} \le 1.
\]

Taking into consideration the inequality (\ref{eq84}) the latter equation can be
written in the following form

\begin{equation}
\label{eq85}
\frac{{1}}{{3}} \le \cos\theta _{0} \le 1.
\end{equation}

Thus, according to the results of this paper in the case when the inequality
(\ref{eq84}) is true the jets of nuclear matter at the collision of heavy nuclei
should to be expected along the directions defined by the condition (\ref{eq85}). In
the case when the inequality (\ref{eq84}) is breaking, the angular distribution of
the outgoing matter should be close to an isotropic one.

Let us come back to the problem collisionless approximation for the
description of dynamics of nuclear matter formed by the
collision of heavy nuclei. The main condition of its applicability (\ref{eq1})
from the point of view of the results of this paper is the inequality
$\omega _{0} \tau _{r} \gg 1$, where $\omega _{0} = skv_{F} $ and $\tau _{r}
$ is a relaxation time. It is necessary also that the characteristic times
of the instabilities' development $\tau _{f} \sim 1 /\gamma $ (where $\gamma $
are the increments found in this paper) should be smaller or have the same
magnitude as the relaxation time $\frac{{1}}{{\gamma} } \lesssim \tau _{r} $. The
latter equation can be written down like $\frac{{\gamma} }{{\omega _{0}
}}\omega _{0} \tau _{r} \gtrsim 1$. Thus, the equations

\[
\omega _{0} \tau _{r} > > 1,
\quad
\frac{{\gamma} }{{\omega _{0}} }\omega _{0} \tau _{r} \gtrsim 1.
\]

\[
\omega _{0}  > > \gamma\gtrsim 1/\tau_r,
\quad
t<<1/\gamma\lesssim\tau_r.
\]

\noindent
are the applicability conditions of the collisionless approximation for the
describing the initial stage of development of the instabilities in the
nuclear matter, formed by the heavy nuclei collisions. Let us remark that
the kinetic equation without taking into account collisions between
particles had been used by some authors for the description of the dynamics
of the matter formed by the heavy nuclei collisions (see in this case
\cite{jR,bBo,mC}, and references therein). As a rule
this equation was written in the mean field approximation. For example the
spinodal fission in the expanding nucleon Fermi-liquid being was
investigated in \cite{mC} on the basis of such kinetic equation. However,
as it is known, the average field approach needs considerable numeric
calculations. The usage of the Fermi-liquid approach with phenomenological
parameters of interaction (Landau amplitudes) for the describing the
dynamics of the nuclear matter allows in many cases to use analytic
treatment without the help of numerical methods.

 As it was repeatedly noted in this paper the application of Fermi-liquid
approach is justified while describing the properties of heavy nuclei. For
this reason the confirmations of the mechanism of origin of instabilities in
the nuclear matter suggested in this paper, one can expect in experiments at
the collisions of heavy fast (but non-relativistic) nuclei. These might be
the reactions Gd+U or Xe+Sn (INDRA, \cite{bBo,jP}) at
the energies of the incoming nucleus more than 145 MeV per nucleon.

The work is performed under STCU (Grant N 1480)  funding
support.


\end{multicols}

\end{document}